\documentclass[twocolumn,showpacs,preprintnumbers,prl]{revtex4}
\usepackage{dcolumn}

\begin{document}

\title{A realistic device that simulates the non-local PR box without
communication}
\author{Sven Aerts}
\address{\textit{Center Leo Apostel for Interdisciplinary Studies (CLEA) and} \\ 
 \textit{ Foundations of the Exact Sciences (FUND) } \\
\textit{Department of Mathematics, Vrije Universiteit Brussel} \\
\textit{Pleinlaan 2, 1050 Brussels, Belgium} \\
email: \textsf{saerts@vub.ac.be}}

\begin{abstract}
A black box with two input bits and two output bits is called a non-local PR
box, if the $XOR$ of the output bits equals the $AND$ of the input bits. In
a recent article\cite{Cerf et al}, Cerf \textit{et al.} show that Alice and
Bob, using such a PR box, can effectively simulate entanglement without the need of communication. 
We show that an adaptation of a model due to
Dirk Aerts\cite{Aerts Vessels}, yields a realistic simulation of the non-local 
PR box without communication. Because the model is entirely realistic,
it cannot violate relativistic constraints. Like a non-local box, it can be
used to simulate the singlet state coincidence probabilities, but the time to complete the
observation of the outcome will exceed the time it takes a photon to travel
one arm in an EPRB setup.  The model explicitly shows how to produce an outcome that is locally perfectly
random, but nevertheless determines what happens in the other wing of the experiment, without communication taking place between the two wings.
In this sense, it can serve as an accurate metaphor for the mechanism of entanglement.
The model considerably strengthens the claim that no communication is necessary to simulate entanglement. 
\end{abstract}

\pacs{03.65.Ud, 03.67.-a }
\maketitle

In a recent article\cite{Cerf et al}, Cerf, Gisin, Massar and Popescu claim
that no communication is necessary to simulate quantum entanglement.
To support their claim, the authors demonstrate that a single instance of a
\textquotedblleft non-local box\textquotedblright\ suffices to simulate the
outcome probabilities of coincidence measurements on a singlet state. A non-local box is
a virtual device with two inputs and two outputs. Alice and Bob each receive
one input bit and are each to produce an output bit such that the $XOR$ of
the output bits equals the $AND$ of the input bits. This constraint between
the input and output is essential and non-trivial. It is unique up to local
transformations of the bits. The device that realizes this logical operation
is called the \textquotedblleft non-local PR box\textquotedblright\ after
the authors that first presented it \cite{PopescuRohrlich}, or simply
\textquotedblleft non-local box \textquotedblright\ (NLB). Before \cite{Cerf et al} appeared,  Toner and Bacon 
showed \cite{TonerBacon} that a single bit of classical communication is
sufficient for the simulation of the outcome probabilities of a singlet
state in an EPRB-type setup. The authors in \cite{Cerf et al} explain that,
because the outcomes of the non-local box could occur in a random fashion
(as long as the constraint is fulfilled), no communication is possible with
this device. Hence the title of the paper: \textquotedblleft Quantum
entanglement can be simulated without communication\textquotedblright .
However, the box is presented as virtual device only and no mechanism is
provided by authors to show \emph{how} the box performs its miraculous task.
The reason being, that a realistic model is generally believed to be ruled out by Bell's theorem. 
Within the setting invoked by Cerf \textit{et al.}, timing is however of no consequence, as they only 
seek to determine the minimal amount of classical information that is sufficient to produce the singlet correlations.
But perhaps the only way to realize such a box, is by means of an internal
signal going from Alice to Bob, and this communication might be concealed by
the box because the box extends all the way from Alice to Bob. Then the
result in \cite{Cerf et al}, although elegant, is still not decisively
stronger than that of Toner and Bacon. We settle this issue here by showing
that such a device can easily be constructed without the need, or indeed the
possibility of communication. The starting point for our investigation, is
that the constraint of the NLB was inspired by the CHSH inequality. In fact,
the NLB violates the CHSH inequality maximally. 
It was however shown \cite{Aerts PhD} as
early as 1982, that one can construct a classical device by means of two
communicating vessels of water, that allows to define measurements with
outcome probabilities that also violate this inequality. Shortly hereafter,
the model was modified to yield a maximal
violation of the CHSH inequality \cite{Aerts Vessels}. It follows that this
classical device might be an instance of such an NLB. Because the communicating vessels model can easily be 
constructed using only classical resources, it cannot violate relativistic constraints and mainly serves the purpose
to provide a structural model of how to generate singlet-like correlations classically.
In a very recent contribution \cite{BroadMethot} have extended the NLB to a 
multi-party setting. Moreover, they show the NLB to be more powerful than quantum correlations  
with respect to certain tasks and express their belief that \textquotedblleft
..\emph{understanding the power of this box, will yield insight into the non-locality
 of quantum mechanics}\textquotedblright . The model we will present
is a step towards such an understanding. The device we propose is
ridiculously simple and is an adaptation of the communicating vessels of
water model. The adaptation is necessary because it can easily be shown that
the original proposal allows Alice to transfer 0.25 bits on average to Bob, 
by merely looking at the outcomes he obtains. This is no longer the case in the model presented here.

The physical model of our non-local box, 
consists of a strectched rubber-band inside a tube that has a length
equal to the distance between Alice and Bob. Alice and Bob each hold on to
one end of the rubber-band. 
The rubber-band has an unstretched length equal to $L$
and can be  yellow or red with equal probability.  When Alice is presented
with input bit $0$, she performs the following measurement. 
She looks at the end of the piece of 
rubber-band she is holding and determines its color. If the rubber-band
is yellow, she returns output bit $1$, if it is red, she will produce a $0$.
When she is given input bit $1$, her operation is slightly more involved.
She will then first count to three to give Bob the time to perform the color
measurement in case he was asked to. Next she will pull the rubber-band. She
measures the length of rubber-band she got and determines its color. If she
finds the length exceeds $L/2,$ she sets a local variable $l_{Alice}$ equal
to $1$, if the length is smaller than $L/2,$ she sets $l_{Alice}=0.$ If the
color is yellow, she will set $c_{Alice}=1,$ if it is red, she will set 
$c_{Alice}=0.$ Her output for the experiment is given by 
$NOT(XOR(l_{Alice},c_{Alice}))$. The $NOT\circ XOR$ (sometimes called XNOR)
operator has a simple interpretation as a correlation indicator. Like a Boolean Kronecker delta, it is equal to one when the two
arguments are the same and equal to zero when its arguments differ. Bob's
experiments are exactly the same, but of course performed on his side of the
experiment. The rubber-band is assumed to be almost maximally stretched in this tube,
so when Alice and Bob both pull the rubber-band, it will break with a uniform
probability, i.e. it will break somewhere in an interval of the rubber-band with a probability
that is proportional only to the length of this interval. If, however, only
one of them pulls the rubber-band, we assume that he/she receives the entire
rubber-band.  This task can be performed by Alice and Bob if each of them
holds the rubber-band just hard enough so the rubber-band won't jump out of their hands. 
Alternatively, this task could be fully automated so that the whole becomes a black box. 
The experiment is finalized as follows.
If, three counts after the initiation of the measurement, neither Alice, nor Bob has started to pull the rubber-band, a mechanism in the middle of the tube
automatically sucks the rubber-band in the tube and out of their hands, and a new rubber-band is put in place. 
This procedure ensures that Alice cannot know whether it was Bob or the mechanism that pulled the rubber-band. 
This concludes the description of the experiment.

The verification that this is an implementation of a non-local box is
somewhat tedious but straightforward.

Suppose Alice and Bob are both given input bit zero, then they both have to
verify whether the rubber-band is yellow, and both will return a $1$ when it is
yellow, or both will return a $0$ when it is red. If we denote the input
bits Alice and Bob receive as a couple $(0, 0),$ then the output bits we get
back, are either $(1, 1)$ or $(0, 0)$. We easily verify that $AND$ $(0, 0)=$ $0=
$ $XOR$ $(1, 1)=$ $XOR$ $(0, 0),$ so that this case fulfills the NLB
constraint.

If Alice and Bob are both given input bit $1,$ they will both count to three
and start pulling the rubber-band. The rubber-band will eventually break somewhere, leaving
either Alice with a piece that is smaller than $L/2$, and Bob a piece that
is larger, or vice versa.\ This means that when they will have set their
local variable $l,$ we will always have that $l_{Alice}=NOT(l_{Bob}).$
Whether the color of the rubber-band is red or yellow, they will both observe
the same color, so that $c_{Alice}=c_{Bob}.$ We obviously have that $%
XOR(l_{Alice},c_{Alice})=XOR(NOT(l_{Bob}),c_{Bob})=$ $%
NOT(XOR(l_{Bob},c_{Bob})).$ From this we see that the output bit Alice gives
is always different from the output bit that Bob gives. Let us verify that
the NLB constraint is fulfilled in this case. The input bits were $(1, 1)$,
so that $AND(1, 1)=1$. Since the output bits are always different, the $XOR$
will yield $1$ too. Please note that, even though the $XOR$ is always one in
this case, the output bits could have been $0$ or $1$ with equal probability.

The last case we need to consider is when Alice receives $0$ and Bob
receives $1$ as input bit. Indeed, we need not verify the constraint when
Alice receives $1$ and Bob receives $0$, because Alice and Bob perform the
same actions when given the same input bit. If Alice receives $0$, she will
output $1$ when the rubber-band is yellow and $0$ when it is red. Bob will count
to three (to give Alice the time to determine the color), and then pulls the
rubber-band to determine its color and length. Suppose Alice found the color was
yellow, then Bob will also determine it is yellow: $c_{Alice}=c_{Bob}=1$.
The model assumes that, if Bob was the only one to pull the rubber-band, he
always receives the entire rubber-band, hence the length of the rubber-band he
measures, will always be greater than $L/2$: $l_{Bob}=1$.\ His outcome is
then $NOT(XOR(1, 1))=1$. So when the input bits are $(0, 1)$, and the rubber-band
was yellow, the output will be $(1,1)$, so that $AND(0, 1)=0=XOR(1, 1),$ which
satisfies the NLB constraint. If however the color of the rubber-band was red,
then Alice outputs a $0$ and Bob finds the color was red too: 
$c_{Alice}=c_{Bob}=0$. Bob once more determines the length to be greater than 
$L/2$: $l_{Bob}=1.$ So Bob gives the output $NOT(XOR(0, 1))=0$. The input in
this case was $(0, 1)$ and the output was $(0, 0)$ and we find once more that 
$AND(0, 1)=0=XOR(0, 0).$ We summarize. For three out of four of the possible
input bit combinations ($(0, 0), (0, 1)$ and $(1, 0)$), the two output bits are the same and yield either 
$(0, 0)$ or $(1, 1)$ with equal probability. It is only when Alice and Bob pull
the rubber-band together, that they break the rubber-band and cause something to
happen `non-locally': the input $(1,1)$ then yields the possible outcomes 
$(0, 1)$ or $(1, 0)$ with equal probability. This concludes the verification of
the NLB constraint for the rubber-band model.

We see that the NLB constraint is fulfilled so that a simple rubber-band inside a tube can be
used to implement it.\ Both Alice and Bob perform only local actions and the
measurement chosen to be performed on one wing of the EPRB experiment, 
\emph{does not\ in itself affect} the possible local outcome on the other wing. To
see this, let us verify this for the various questions. In case the question
was about the color of the rubber-band, the result was predetermined when the
rubber-band was placed in the tube, so that no influence was exercised at all.
If Alice is measuring the length, but Bob is measuring the color (or vice
versa), then the results are predetermined as well. If they are both
measuring the length of the rubber-band, then the results are not predetermined, 
but it is still only when she has obtained \emph{a definite outcome}, that she can conjecture the outcome Bob
has obtained, given that Bob has measured the length too. As long as no
definite outcome has been obtained, we can clearly still not determine what
has happened on the other side, because for each experiment, we will still
have that, locally, every outcome is equally likely. 
We cannot know the outcome before we have broken the rubber-band and we simply do not know where
the rubber-band will break. Clearly, no communication is possible between Alice
and Bob. But neither, and this is more important but also more subtle, 
can we speak of some internal, hidden signal in the NLB going from Alice to Bob.
What really happens in this simple model, is that Alice and Bob  \emph{together} create opposite outcomes. 
This is why neither Alice, nor Bob, can control the outcome, and yet the two outcomes they obtain are perfectly anti-correlated.
This seems to be in accordane with what happens in the quantum case.
Of course, one can modify this setup so that Alice and Bob can use the device to communicate, but because the model is
fully causal and realistic, it obeys the constraints of special relativity.
In this sense our simulation of the non-local box is, of course, local. So the rubber-band model can
reproduce the correct coincidence probabilities of the EPRB experiment. The
rubber-band might even be elongated at a speed close to the speed of light
before it suddenly stops and Alice and Bob perform their measurements. But
once we start the observation, and Alice and Bob are given input bit 1, the
rubber-band will break in two pieces that need to travel the space between the
breaking point and Alice and Bob. Because of relativistic constraints on
the particles that constitute the rubber-band, the time it takes to complete the
measurement will therefore exceed the travelling time of the photon in a
typical EPRB experiment. It is plausible that this restriction
allows a refutation of this model for these experiments, but this is of
no concern to us here. The model only serves to illustrate that there is a
class of realistic phenomena that allows the production of quantum
correlations without communication. If one cares to take this model
seriously as a metaphor, we are led to the conclusion that the cause of the
entanglement is not so much that there is an information flow from Alice to Bob (or vice versa),
but rather that Alice and Bob are able to \emph{co-create information between them}.
That this is physically feasible, seems to point out that even single entities cannot always be confined to a small
spatial region, and that measurements can change the state. One of the structurally most intriguing 
aspects of quantum correlations, is that the \emph{outcome} of a given measurement performed in one
wing of the EPRB experiment, seems to determine the outcome in the other wing,
even though the outcome itself is, locally speaking, completely random. The model captures this peculiar effect quite 
naturally. In conclusion, we believe to have shown that the claim by Cerf \textit{et al.}
that entanglement can be simulated without communication is strengthened by
providing a realistic model that does just that. \\ 
Acknowledgement: The author thanks Anne Broadbent for kindly pointing out a mistake in the abstract and references.


\begin{thebibliography}{9}

\bibitem{Cerf et al} N.J. Cerf, N. Gisin, S. Massar and S. Popescu. Quantum
entanglement can be simulated without communication, quant-ph/0410027, (2004).

\bibitem{Aerts Vessels} D. Aerts. Example of a macroscopical situation that
violates Bell inequalities. Lettere al Nuovo Cimento, \textbf{34}, pp.
107-111, (1982). This article can be downloaded at: 
{\scriptsize www.vub.ac.be/CLEA/aerts/publications/1982ExMacViolBell.pdf}

\bibitem{PopescuRohrlich} S. Popescu and D. Rohrlich. quant-ph/9709026, (1997).

\bibitem{TonerBacon} B. F. Toner and D. Bacon. Communication Cost of Simulating Bell Correlations, 
Phys Rev Lett. \textbf{91}, 187904, (2003).

\bibitem{Aerts PhD} D. Aerts. The One and the Many: Towards a Unification of
the Quantum and Classical Description of One and Many Physical Entities.
Doctoral dissertation, Free University of Brussels, (1982).

\bibitem{BroadMethot} A. Broadbent and Andr\'{e} Allan M\'{e}thot. On the power
of non-local boxes, quant-ph/0504136, (2005).
\end{thebibliography}
\end{document}